\documentclass[fleqn,10pt]{wlscirep}
\usepackage{geometry}
\usepackage{etoolbox}
\usepackage{graphicx}
\usepackage{color}
\usepackage{graphicx}
\usepackage{amsmath}
\usepackage{capt-of}
\usepackage{caption}
\usepackage{slashbox}

\title{Microwave-induced orbital angular momentum transfer}

\author[1]{Zahra Amini Sabegh}
\author[1]{Mohammad Ali Maleki}
\author[1,*]{Mohammad Mahmoudi}
\affil[1]{Department of Physics, University of Zanjan, University Blvd., 45371-38791, Zanjan, Iran}
\affil[*]{mahmoudi@znu.ac.ir}

\begin{abstract}
The microwave-induced orbital angular momentum (OAM) transfer from a Laguerre-Gaussian (LG) beam to a weak plane-wave is studied in a closed-loop four-level ladder-type atomic system. The analytical investigation shows that the generated fourth field is a LG beam with the same OAM of the applied LG field. Moreover, the microwave-induced subluminal generated  pulse can be switched to the superluminal one only by changing the relative phase of applied fields. It is shown that the OAM transfer in subluminal regime is accompanied by a slightly absorption, however, it switches to the slightly gain in superluminal regime. The transfer of light's OAM and control of the group velocity of generated pulse can prepare a high-dimensional Hilbert space which has a major role in quantum communication and information processing.
\end{abstract}
\begin{document}
\flushbottom
\maketitle

\thispagestyle{fancy}

In the past three decades, much attention has been paid to the study of the optical phenomena using Laguerre-Gaussian (LG) laser fields which are derived by solving the Helmholtz equation in the cylindrical coordinates. The intensity profile of the LG beam has a doughnut-shaped pattern whose origin is the phase singularity and the azimuthal component of the beam's Poynting vector. It is well known that a light beam may have both the orbital angular momentum (OAM) due to its helical wavefront and the spin angular momentum associated with the circular polarization \cite{book2011,today2004}. Allen \textit{et al.}, for the first time, introduced a well-defined OAM for a LG laser light by measuring the induced mechanical torque on the suspended cylindrical lenses due to the OAM transfer \cite{PRA1992}. The LG modes can be practically generated in several common methods such as forked diffraction gratings \cite{forked1990}, computer-generated holograms \cite{holo1992}, cylindrical lens mode converters \cite{HG1993}, spiral phase plates \cite{plate1994}, and spatial light modulators (SLMs) \cite{SLM2007}. Light beams with OAM were strongly used in fundamental physical subjects like nonlinear optical phenomena, angular momentum conservation \cite{PRA1996}, quantum effects \cite{zeilinger2001}, nano optics \cite{science2011}, imaging \cite{imaging2009}, and quantum communications \cite{exp2004}.

In the recent years, the LG beams and their features have been extensively investigated in many aspects. The effect of the LG intensity profile on the optical spectrum linewidth of electromagnetically induced transparency (EIT) \cite{Hanle2010,Sapam2011,Akin2014}, multi-photon resonance phenomena \cite{Kazemi2016}, optical trapping potential \cite{Kazemi2017}, and electromagnetically induced focusing \cite{Amini} were studied in different atomic media. The spatially dependent EIT has been demonstrated by probing cold rubidium atoms with optical vortex light \cite{PRL2015}. The reflection properties of the OAM of the LG beams at conventional and phase-conjugation mirrors were experimentally studied \cite{exp2009}. In an experimental work, phase and interference characteristics of optical vortex beams were investigated which were in good agreement with the theoretical prediction \cite{JOSAA2008}. Generation of the optical vortex beams from a non-vortex beam occurred by a wave mixing process in a nonlinear photonic crystal \cite{exp2007}. Several nonlinear optical processes have been studied in different quantum systems using LG beams \cite{PRA2017} such as the second-harmonic generation \cite{PRA1996,PRA1997}, sum frequency generation \cite{JOSAB2015}, and four-wave mixing (FWM) \cite{PRL2012,lett2004,PRB2015,Hamedi2018}. The spatially structured optical transparency is theoretically investigated in a five-level combined tripod and $\Lambda$ atom-light coupling scheme \cite{exp2018}. Mahmoudi \textit{et al.} have reported the spatially dependent atom-photon entanglement using LG laser beams. It was shown that the atom-photon entanglement can be controlled by OAM of light in closed-loop atomic systems \cite{Srep2018}. The group velocity of the LG beams in the free space has been also studied. Recently, Boyd \textit{et al.} have shown that the group velocity of a twisted light depends on the distance across the light propagation direction in vacuum \cite{optica2016}. In another work, the subluminal group velocity of the LG beam was reported and it was shown that an optical vortex beam is dispersed even in the free space \cite{Srep2016}.

On the other hand, the OAM transfer from a pump field to both probe and generated fields at the FWM frequency was suggested in a two-level system \cite{PRA2001}. In the following, it was shown that the OAM transfer from two vortex control fields to a probe field occurs in both four-level tripod-type and double tripod-type atomic systems \cite{PRA2011,PRA2013}. Recently, the transfer of the optical vortex in four-level double-$\Lambda$ EIT scheme, through a FWM process, has been reported only by one vortex control field \cite{Hamedi2018}.

In this manuscript, we are going to introduce a model in which the microwave field induces an OAM transfer from strong coupling LG filed to the generated fourth field (GFF). We consider a closed-loop four-level ladder-type atomic system and investigate the effect of the different parameters on the OAM transfer, as well as the intensity and phase profiles. In addition, we investigate the spatially dependent dispersion and absorption of the GFF and obtain an analytical expression to explain the group velocity behavior of the GFF. It is shown that a weak microwave probe field transfers the OAM of a strong coupling LG field to the GFF. The OAM transfer has been already reported just with the EIT window in the absorption doublet. Here, we find new type of the OAM exchange which is ensured by the EIT window in the gain doublet of the GFF. Due to the different scattering precoces of the applied fields in the closed-loop quantum systems, the OAM can be transferred from an applied LG field to another laser field and can be controlled by the relative phase of the applied fields. The GFF is a LG field and can superluminally propagate passing through the atomic vapor cell, for the special set of parameters. Thus, our model proposes a simple method to transfer the OAM from a LG beam to a weak plane-wave using the microwave field in the gain doublet EIT window. Moreover, the slope of dispersion of microwave-induced GFF switches from positive to negative only by changing the relative phase of the applied fields. It is worth to note that, because of the degree of freedom for OAM, the OAM transfer and switching of GFF group velocity can be used for transfer, storage, and processing of high-dimensional optical information.

\section*{Theoretical framework}

\begin{figure}[htbp]
\centering
  \includegraphics[width=4.0cm]{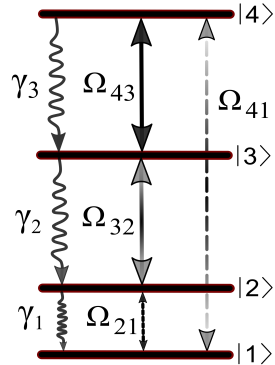}\\
  \caption{\small Schematics of the closed-loop four-level ladder-type atomic system which can be established in $^{87}Rb$ applying a weak planar microwave field, $\Omega_{21}$, strong coupling LG field, $\Omega_{32}$, strong coupling planar field, $\Omega_{43}$, and weak planar field, $\Omega_{41}$. The spontaneous emission rates are indicated by $\gamma_{1}$, $\gamma_{2}$ and $\gamma_{3}$.}\label{f1}
\end{figure}

The proposed four-level ladder-type atomic system consisting of four energy states is illustrated in Fig. \ref{f1}. We consider an ensemble of $^{87}Rb$ atoms as a realistic example with $|1\rangle=|5~^{2}S_{1/2},F=1\rangle$, $|2\rangle=|5~^{2}S_{1/2},F=2\rangle$, $|3\rangle=|5~^{2}P_{3/2}\rangle$, and $|4\rangle=|7~^{2}S_{1/2}\rangle$. The transition $|i\rangle\leftrightarrow|j\rangle$ is driven by an external field with frequency $\omega_{ij}$ and Rabi frequency $\Omega_{ij}=\vec{\mu}_{ij}.\vec{E}_{ij}/\hbar$ ({$i,j\in {1,...,4}$}). Here, $\mu_{ij}$ and $E_{ij}$ are the induced dipole moment of the transition $|i\rangle\leftrightarrow|j\rangle$ and the amplitude of the applied field, respectively. A planar microwave field is applied to the $|1\rangle\leftrightarrow|2\rangle$ transition as the first weak probe field. The transition $|2\rangle\leftrightarrow|3\rangle$ is excited by a strong coupling LG field which in the cylindrical coordinates has the form
\begin{eqnarray}\label{e1}
    E_{32}(r,\varphi)&=&E_{0_{32}}\frac{1}{\sqrt{|l|!}}(\frac{\sqrt{2}r}{w_{LG}})^{|l|}~e^{-r^{2}/w_{LG}^{2}}~
    e^{i[\frac{n\omega_{32}}{c}(z+\frac{r^2z}{2(z^2+z_R^2)^2})-(|l|+1)tan^{-1}(z/z_R)+l\varphi]},
\end{eqnarray}
where the field strength, LG beam waist, OAM, refractive index of medium, light frequency, velocity of light in vacuum, and propagation direction of light are denoted by $E_{0_{32}}$, $w_{LG}$, $l$, $n$, $\omega_{32}$, $c$, and $z$, respectively, and $z_R=n\omega_{32} w_{LG}^2/2c$ being the Rayleigh range. Two planar fields are exerted to the $|3\rangle\leftrightarrow|4\rangle$ and $|1\rangle\leftrightarrow|4\rangle$ transitions as the second strong coupling and weak probe fields.
The spontaneous emission rates for the transitions $|1\rangle\leftrightarrow|2\rangle$, $|2\rangle\leftrightarrow|3\rangle$ and $|3\rangle\leftrightarrow|4\rangle$ are indicated by $\gamma_{1}$, $\gamma_{2}$ and $\gamma_{3}$.
In the rotating-wave and electric-dipole approximations, the time evolution of this system can be described by the density matrix equations of motion which stand for
\begin{eqnarray}\label{e2}
  \dot{\rho}_{11}&=&\gamma_{1}\rho_{22}+i\Omega^{\ast}_{21}\rho_{21}-i\Omega_{21}\rho_{12}+i\Omega^{\ast}_{41}\rho_{41}
  -i\Omega_{41}\rho_{14},\nonumber\\
  \dot{\rho}_{22}&=&-\gamma_{1}\rho_{22}+\gamma_{2}\rho_{33}+i\Omega_{21}\rho_{12}-i\Omega^{\ast}_{21}\rho_{21}
  +i\Omega^{\ast}_{32}\rho_{32}-i\Omega_{32}\rho_{23},\nonumber\\
  \dot{\rho}_{33}&=&-\gamma_{2}\rho_{33}+\gamma_{3}\rho_{44}+i\Omega_{32}\rho_{23}-i\Omega^{\ast}_{32}\rho_{32}
  +i\Omega^{\ast}_{43}e^{-i\phi_0}\rho_{43}-i\Omega_{43}e^{i\phi_0}\rho_{34},\nonumber\\
  \dot{\rho}_{12}&=&-(i\Delta_{21}+\gamma_{1}/2)\rho_{12}-i\Omega_{32}\rho_{13}+i\Omega^{\ast}_{21}(\rho_{22}-\rho_{11})
  +i\Omega^{\ast}_{41}\rho_{42},\nonumber\\
  \dot{\rho}_{13}&=&-[i(\Delta_{21}+\Delta_{32})+\gamma_{2}/2]\rho_{13}+i\Omega^{\ast}_{21}\rho_{23}-i\Omega^{\ast}_{32}\rho_{12}
  -i\Omega_{43}e^{i\phi_0}\rho_{14}+i\Omega^{\ast}_{41}\rho_{43},\nonumber\\
  \dot{\rho}_{14}&=&-[i\Delta_{41}+\gamma_{3}/2]\rho_{14}-i\Omega^{\ast}_{43}e^{-i\phi_0}\rho_{13}+i\Omega^{\ast}_{21}\rho_{24}
  +i\Omega^{\ast}_{41}(\rho_{44}-\rho_{11}),\nonumber\\
  \dot{\rho}_{23}&=&-[i\Delta_{32}+(\gamma_{1}+\gamma_{2})/2]\rho_{23}+i\Omega_{21}\rho_{13}-i\Omega_{43}e^{i\phi_0}\rho_{24}
  +i\Omega^{\ast}_{32}(\rho_{33}-\rho_{22}),\nonumber\\
  \dot{\rho}_{24}&=&-[i(\Delta_{32}+\Delta_{43})+(\gamma_{1}+\gamma_{3})/2]\rho_{24}+i\Omega^{\ast}_{32}\rho_{34}
  +i\Omega_{21}\rho_{14}-i\Omega^{\ast}_{43}e^{-i\phi_0}\rho_{23}-i\Omega^{\ast}_{41}\rho_{21},\nonumber\\
  \dot{\rho}_{34}&=&-[i\Delta_{43}+(\gamma_{2}+\gamma_{3})/2]\rho_{34}+i\Omega_{32}\rho_{24}
  +i\Omega^{\ast}_{43}e^{-i\phi_0}(\rho_{44}-\rho_{33})-i\Omega^{\ast}_{41}\rho_{31},\nonumber\\
  \dot{\rho}_{44}&=&-(\dot{\rho}_{11}+\dot{\rho}_{22}+\dot{\rho}_{33}),
\end{eqnarray}
where $\Delta_{ij}=\omega_{ij}-\bar{\omega}_{ij}$ defines frequency detuning between the applied laser field and $|i\rangle\leftrightarrow|j\rangle$ transition and $\phi_0$ is the relative phase of the applied fields.

The generation of a fourth field with the frequency $\omega_{41}=\omega_{21}+\omega_{32}+\omega_{43}$, known as FWM process, is a common phenomenon in which energy is completely conserved. Then, we study the response of the medium to the applied fields using the susceptibility of the $|1\rangle\leftrightarrow|4\rangle$ transition which is proportional to the coherence term of the transition, $\rho_{41}$. The assumption is that the positive imaginary part of the susceptibility is supposed to be the absorption response of the atomic system while its negative value indicates the gain. Note that the dispersion is determined by the real part of the $\rho_{41}$ so that the positive (negative) slope of dispersion indicates the normal (anomalous) dispersion. In the special conditions, $\gamma_{1}=\gamma_{2}=\gamma$, $\gamma_{3}=0$, $\Delta_{32}=\Delta_{43}=0$, and $\Delta_{41}=\Delta_{21}=\delta$, we obtain the following analytical expressions
\begin{eqnarray}\label{e3}
  \rho_{21}&=&\frac{2[\Omega_{21}(i\gamma\delta-2|\Omega_{43}|^{2})+2\Omega_{41}\Omega_{32}^{*}\Omega_{43}^{*}e^{-i\phi_0}]}
  {\gamma^{2}\delta+2i\gamma|\Omega_{43}|^{2}+4\delta(|\Omega_{32}|^{2}+|\Omega_{43}|^{2})},\nonumber\\
  \rho_{41}&=&\frac{4\Omega_{21}\Omega_{32}\Omega_{43}e^{i\phi_0}+(4i\gamma\delta-\gamma^{2}-4|\Omega_{32}|^{2})\Omega_{41}}{\gamma^{2}\delta+2i\gamma
  |\Omega_{43}|^{2}+4\delta(|\Omega_{32}|^{2}+|\Omega_{43}|^{2})},
\end{eqnarray}
for the coherence terms of the probe transitions, in the steady state. Now, we are going to find the amplitude of the GFF using the Maxwell equations. In the slowly varying envelope approximation and assumption of time-independent probe fields the Maxwell equations for two weak probe fields, propagating in $z$ direction, read to
\begin{eqnarray}\label{e4}
  \frac{\partial\Omega_{21}(z)}{\partial z}&=&i\frac{\alpha\gamma}{2L}\rho_{21},\nonumber\\
  \frac{\partial\Omega_{41}(z)}{\partial z}&=&i\frac{\alpha\gamma}{2L}\rho_{41},
\end{eqnarray}
 where $\alpha$ and $L$ are the optical depth of the probe fields and atomic vapor cell length, respectively \cite{Hamedi2018}. Considering the initial conditions for the probe fields amplitudes at the atomic medium entrance as $\Omega_{41}(z=0)=0$ and $\Omega_{21}(z=0)=\Omega_{21}(0)$, and $\delta=0$, simultaneous solving of equations (\ref{e3}) and (\ref{e4}) leads to an explicit term for the Rabi frequency of the GFF as
\begin{eqnarray}\label{e5}
  \Omega_{41}(z)&=&\frac{8\Omega_{21}(0)\Omega_{32}\Omega_{43}e^{i\phi_0}}{A}~e^{-\alpha z[\gamma^{2}+4(|\Omega_{32}|^{2}+|\Omega_{43}|^{2})]/8L|\Omega_{43}|^{2}}~sinh(\alpha z A/8L|\Omega_{43}|^{2}),
\end{eqnarray}
in which
\begin{eqnarray*}
  A=\sqrt{16|\Omega_{32}|^{4}+(\gamma^{2}-4|\Omega_{43}|^{2})^{2}+8|\Omega_{32}|^{2}(\gamma^{2}+4|\Omega_{43}|^{2})}.
\end{eqnarray*}
According to equation (\ref{e5}), the phase factor of the strong coupling LG field transfers to the GFF, in the presence of two other planar fields. The planar microwave field has a key role in the OAM transfer so that there will not be the GFF in the absence of the microwave field . On the other hand, we would like to investigate the group velocity of the generated pulse inside the atomic vapor cell. In a dispersive medium, the different frequency components of the light pulse feel different refractive indices which affect the group velocity of pulse. The group velocity of a pulse is given by $v_{g}=|\partial_\omega\nabla\Phi|^{-1}$, where $\Phi$ stands for the phase factor of the propagating pulse. In the proposed atomic model, the group velocity of the GFF at maximum intensity ring, $r_{max}=w_{LG}\sqrt{|l|/2}$, is given by
\begin{eqnarray}\label{e6}
  v_{g}(z)=\frac{c(4c^2z^2+w_{LG}^4\omega^2n^2)^{3}}{(n+\omega \frac{\partial n}{\partial \omega})B},
\end{eqnarray}
with
\begin{eqnarray*}
  B&=&\{16c^4[-16c^4r_{max}z^5+r_{max}w_{LG}^8z\omega^4n^4]^2+[-32c^6z^4(r_{max}^2+(l+1)w_{LG}^2-2z^2)\nonumber\\
  &+&48c^4w_{LG}^4z^2\omega^2n^2(r_{max}^2+z^2)+2c^2w_{LG}^8\omega^4n^4(-r_{max}^2+(l+1)w_{LG}^2+6z^2)+w_{LG}^{12}\omega^6n^6]^2\}^{1/2}.
\end{eqnarray*}
  where $\omega$ is the GFF frequency. Here, the refractive index can be approximated in terms of the real part of the susceptibility as $n\approx1+\chi'/2$. According to equations (\ref{e3}) and (\ref{e6}), the group velocity of a pulse can exceed the velocity of light in vacuum, leading to the superluminal light propagation.

\section*{Results and discussions}

\begin{figure}[htbp]
\centering
  \includegraphics[width=0.45\textwidth]{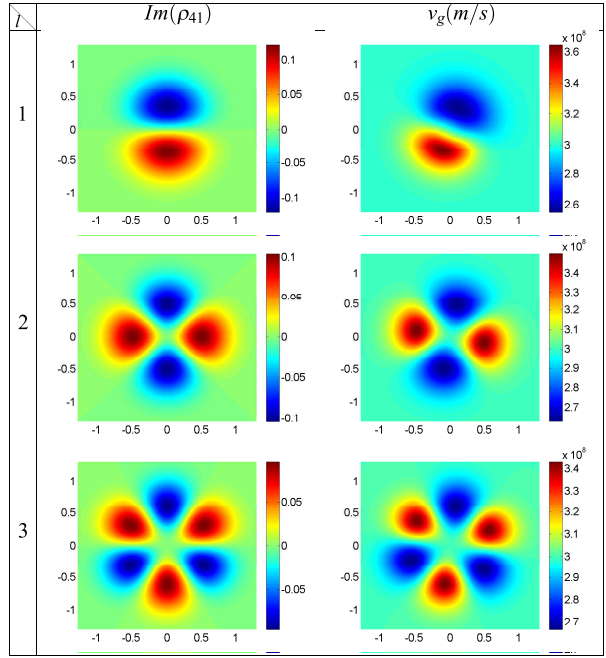}\\
  \caption{\small Imaginary part of $\rho_{41}$ and GFF group velocity profiles as a function of $x$ and $y$ for different modes of the strong coupling LG field with $l=1,2,3$ at the entrance of the atomic medium, $z=0$. The horizontal, $x$, and vertical, $y$, axes are taken in $mm$. Used parameters are $\Delta_{32}=\Delta_{43}=0$, $\Delta_{41}=\Delta_{21}=\delta$ which is considered zero for left column, $\omega=788.7 THz$, $\phi_0=0$, $w_{LG}=0.5 mm$, $\Omega_{21}(0)=0.1\gamma$ and $\Omega_{0_{32}}=\Omega_{43}=10\gamma$.}\label{f2}
\end{figure}

\begin{figure}[htbp]
\centering
  \includegraphics[width=0.45\textwidth]{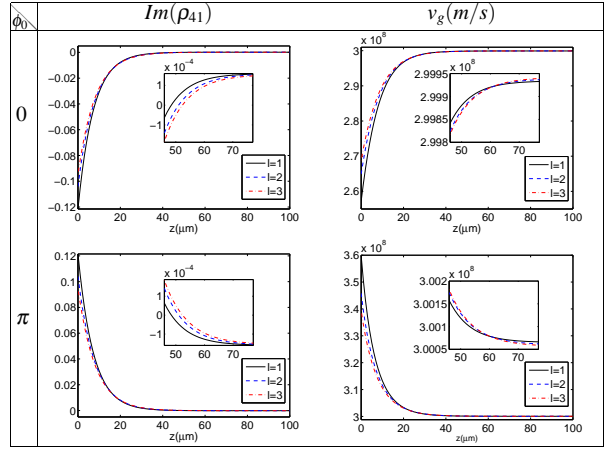}\\
  \caption{\small Imaginary part of $\rho_{41}$ and GFF group velocity behaviors versus of $z$ for different modes of the strong coupling LG field with $l=1,2,3$ and two selected relative phases $\phi_0=0,\pi$, at $r_{max}$. The other used parameters are the same as in Fig. \ref{f2}.}\label{f3}
\end{figure}

In this section, we present the results of our analytical calculations describing the OAM transfer from the strong coupling LG beam to a weak plane-wave via a weak planar microwave field. Moreover, we study the group velocity of the generated pulse, based on the equation (\ref{e6}), in the proposed atomic model. So, we focus on the coherence term $\rho_{41}$, to investigate the spatially dependent absorption, intensity and phase profiles and group velocity of the GFF. All Rabi frequencies are scaled by $\gamma=2\pi\times6 MHz$. Firstly, in Fig. \ref{f2}, we plot the imaginary part of $\rho_{41}$ (left column) and GFF group velocity (right column) profiles as a function of $x$ and $y$ for different modes of the strong coupling LG field with $l=1,2,3$ at the entrance of the atomic medium, $z=0$. The horizontal, $x$, and vertical, $y$, axes are taken in $mm$. Used parameters are $\Delta_{32}=\Delta_{43}=0$, $\Delta_{41}=\Delta_{21}=\delta$ which is considered zero for the left column, and $\omega=788.7 THz$. Other characteristics of the applied fields are chosen to be $w_{LG}=0.5 mm$, $\Omega_{21}(0)=0.1\gamma$, and $\Omega_{0_{32}}=\Omega_{43}=10\gamma$. The spatially-dependent imaginary part of $\rho_{41}$ profile shows a petal-like pattern with $2l$ number of petals. Right column of Fig. \ref{f2} shows that there are some regions of high absorption (gain) with superluminal (subluminal) group velocity. On the other hand, a small rotation in group velocity petal-like patterns implies the gain  or absorption doublet for the GFF in the atomic system. This result persuaded us to investigate the group velocity behavior of the GFF and imaginary part of $\rho_{41}$, further in its propagation direction inside the atomic system.
Figure \ref{f3} shows the imaginary part of $\rho_{41}$ (left column) and group velocity behavior of the GFF (right column) versus $z$ for different modes of the strong coupling LG field, $l=1,2,3$, in two relative phases, $\phi_0=0,\pi$, at $r_{max}$. The other used parameters are the same as in Fig. \ref{f2}. It is found that, by propagating the GFF in $z$ direction, the negative imaginary part of $\rho_{41}$ switches to the EIT with a small absorption which is accompanied by subluminal group velocity. Moreover, a switching between the superluminal absorption and the superluminal gain doublet happens with a negligible gain for $\phi_0=\pi$.
Note that the GFF attenuates during its propagation in the presence of the absorption and the OAM cannot effectively transfer from the strong coupling LG beam to the GFF. However, such disadvantage can be removed by changing the relative phase to $\phi_0=\pi$. Then, the OAM transfer can be simply controlled by changing the relative phase of the applied fields in the closed-loop proposed atomic system.

\begin{figure}[htbp]
\centering
  \includegraphics[width=0.45\textwidth]{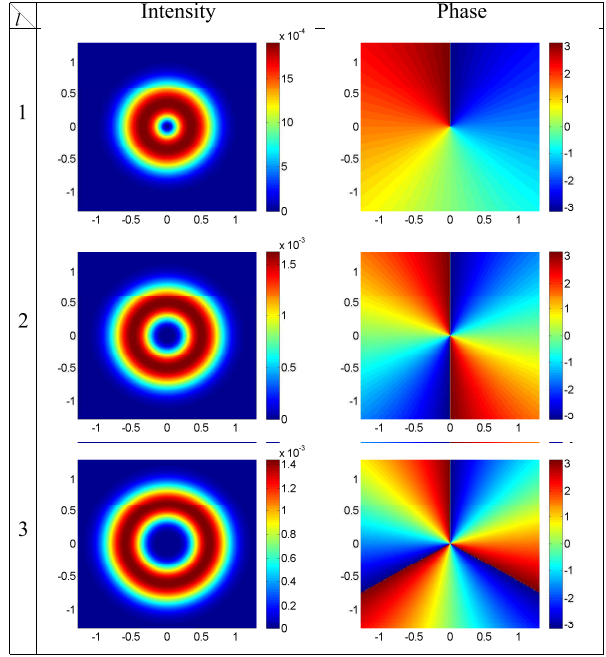}\\
  \caption{\small Intensity and phase profiles of the GFF at the end of atomic vapor cell, $z=L=100\mu m$, as a function of $x$ and $y$ for the first three modes of the strong coupling LG field, $l=1,2,3$, $\alpha=10$, $\delta=0$, and $\phi_0=\pi$, with the same parameters as used in Fig \ref{f2}.}\label{f4}
\end{figure}
\begin{figure}[htbp]
\centering
  \includegraphics[width=0.45\textwidth]{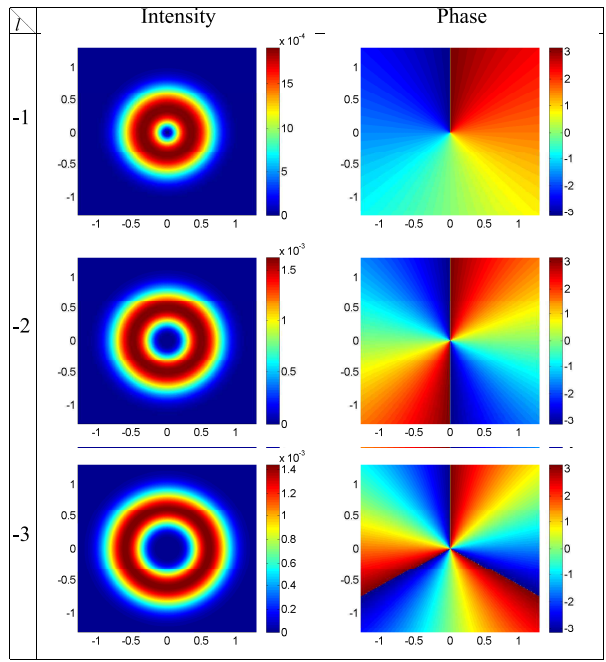}\\
  \caption{\small Intensity and phase profiles of the GFF at $z=L=100\mu m$ versus $x$ and $y$ for different modes of the strong coupling LG field with the negative OAMs, $l=-1,-2,-3$, in which other parameters are same as in Fig. \ref{f4}.}\label{f5}
\end{figure}

Let us now, to study the optical properties of the GFF for $\phi_0=\pi$. Using equation (\ref{e5}), the intensity and phase profiles of the GFF as a function of $x$ and $y$ are plotted at the end of the atomic vapor cell, $z=L=100\mu m$, for the first three modes of the strong coupling LG field, $l=1,2,3$, $\alpha=10$, $\delta=0$, and $\phi_0=\pi$ in Fig. \ref{f4}, with the same parameters as used in Fig \ref{f2}. As expected from equation (\ref{e5}), the GFF is a LG field with the OAM of the strong coupling LG field. The OAM transfer for the negative OAMs, $l=-1,-2,-3$, is illustrated in Fig. \ref{f5}, in which other parameters are same as in Fig. \ref{f4}. Overall, our model introduces a microwave-induced OAM transfer between the light beams via FWM process which is accompanied by the gain-assisted superluminal pulse propagation.

\section*{Conclusion}

In conclusion, we have theoretically investigated the microwave-induced OAM transfer and the group velocity of the GFF in a four-level ladder-type atomic system. It has been shown that the weak planar microwave field induces an OAM transfer from a strong coupling LG field to the GFF via FWM process. Moreover, we obtained an analytical expression to describe the GFF group velocity for different values of the OAMs. In addition, it has been found that the relative phase of the applied fields has a crucial role in controlling the OAM transfer as well as the group velocity of the GFF. We found a proper relative phase which switches the subluminal to the superluminal GFF propagation. The OAM transfer accompanied by the gain-assisted superluminal pulse propagation can be used in the development of optical communication, quantum information processing, and data transfer in high-dimensional Hilbert space.


\section*{Legends}

\setlength{\parindent}{0ex}

Fig. \ref{f1}. Schematics of the closed-loop four-level ladder-type atomic system which can be established in $^{87}Rb$ applying a weak planar microwave field, $\Omega_{21}$, strong coupling LG field, $\Omega_{32}$, strong coupling planar field, $\Omega_{43}$, and weak planar field, $\Omega_{41}$. The spontaneous emission rates are indicated by $\gamma_{1}$, $\gamma_{2}$ and $\gamma_{3}$.

Fig. \ref{f2}. Imaginary part of $\rho_{41}$ and GFF group velocity profiles as a function of $x$ and $y$ for different modes of the strong coupling LG field with $l=1,2,3$ at the entrance of the atomic medium, $z=0$. The horizontal, $x$, and vertical, $y$, axes are taken in $mm$. Used parameters are $\Delta_{32}=\Delta_{43}=0$, $\Delta_{41}=\Delta_{21}=\delta$ which is considered zero for left column, $\omega=788.7 THz$, $\phi_0=0$, $w_{LG}=0.5 mm$, $\Omega_{21}(0)=0.1\gamma$ and $\Omega_{0_{32}}=\Omega_{43}=10\gamma$.

Fig. \ref{f3}. Imaginary part of $\rho_{41}$ and GFF group velocity behaviors versus of $z$ for different modes of the strong coupling LG field with $l=1,2,3$ and two selected relative phases $\phi_0=0,\pi$, at $r_{max}$. The other used parameters are the same as in Fig. \ref{f2}.

Fig. \ref{f4}. Intensity and phase profiles of the GFF at the end of atomic vapor cell, $z=L=100\mu m$, as a function of $x$ and $y$ for the first three modes of the strong coupling LG field, $l=1,2,3$, $\alpha=10$, $\delta=0$, and $\phi_0=\pi$, with the same parameters as used in Fig \ref{f2}.

Fig. \ref{f5}. Intensity and phase profiles of the GFF at $z=L=100\mu m$ versus $x$ and $y$ for different modes of the strong coupling LG field with the negative OAMs, $l=-1,-2,-3$, in which other parameters are same as in Fig. \ref{f4}.

\section*{Author contributions}
M.M. proposed the idea of the microwave-induced OAM transfer. All authors developed the research conceptions, analysed, and discussed the obtained results. Z.A.S. and M.A.M performed the calculations and wrote the paper with major input from M.M.

\section*{Additional information}
\textbf{Competing financial interests:} The authors declare no competing financial and non-financial interests.

\end{document}